\documentclass[a4paper,10pt,pra,aps,twocolumn]{revtex4}
\usepackage{amsmath,amssymb,graphicx}
\usepackage{braket, datetime}
\usepackage[usenames,dvipsnames]{color}

\newcommand{\figref}[1]{Fig.~\ref{#1}}
\newcommand{\eqnref}[1]{Eqn.~(\ref{#1})}

\definecolor{DarkRed}{rgb}{0.7,0,0}
\definecolor{DarkBlue}{rgb}{0,0,0.7}
\definecolor{DarkGreen}{rgb}{0,0.5,0}
\newcommand{\tobeignored}[1]{{}}
\newcommand{\creation}{\ensuremath{{\psi}^*}}
\newcommand{\annihilation}{\ensuremath{{\psi}}}

\newcommand{\drr}{{\rm d}x\,{\rm d}y}
\newcommand{\ei}[1]{e^{{\rm i} #1}}
\newcommand{\eim}[1]{e^{-{\rm i} #1}}
\newcommand{\partialderiv}[2]{\frac{\partial #1}{\partial #2}}
\newcommand{\dpsi}{\ensuremath{\delta\annihilation}}
\newcommand{\dpsidag}{\ensuremath{\delta\creation}}
\newcommand{\kom}{\ensuremath{{\bf k},\omega}}
\newcommand{\ek}{\ensuremath{\epsilon_{\bf k}}}
\newcommand{\xik}{\ensuremath{\xi_{\,\bf k}}}
\newcommand{\kxlam}{\ensuremath{(k_x,\lambda)}}

\begin{document}

\title{Interactions in dye-microcavity photon condensates and
the prospects for their observation}
\author{R. A. Nyman}\email[Correspondence should be addressed to
]{r.nyman@imperial.ac.uk} \affiliation{Centre for Cold Matter,
Blackett Laboratory, Imperial College London, Prince Consort Road, SW7
2BW, United Kingdom}

\author{M. H. Szyma\'nska}\altaffiliation{Present address: Department of Physics and Astronomy, University College London, Gower Street, London WC1E 6BT, United Kingdom} \affiliation{Department of Physics,
University of Warwick, Coventry CV4 7AL, United Kingdom}

%\date{\today, \ampmtime}
\begin{abstract}
	We derive the equation of motion for a Bose-Einstein condensate of photons in a dye-microcavity system, starting from Maxwell's equations. Our theory takes into account mirror shape, Kerr-type intensity-dependent refractive index and incoherent pumping and loss. The resulting equation is remarkably similar to the Gross-Pitaevskii equation for exciton-polariton condensates, despite the different microscopic origins. We calculate the incoherent photoluminescence spectrum of the photon condensate which shows the Bogoliubov-type excitations around the mean-field at thermal equilibrium. Both open and closed-system models are presented to account for, respectively dissipation and inhomogeneities. Considering realistic parameters and experimental resolution, we estimate that by observing the angle-resolved spectrum of incoherent photoluminescence it is possible to resolve dimensionless interaction parameters of order $10^{-5}$, two orders of magnitude below current estimates. Thus we expect that this technique will lead to accurate measurements of the interactions in photon condensates. 
\end{abstract}

%PACS NUMBERS (to be updated when webpage is back up. Try also: 67.85.Hj 
%03.75.Hh; 42.50.-p; 67.90.+z

\maketitle

%\section{Introduction}

Bose-Einstein condensation is usually thought of as a low-temperature phenomenon. However, Klaers \textit{et al} made the first room-temperature Bose-Einstein condensate (BEC) \cite{Klaers10b} by confining photons in a cavity filled with fluorescent dye to provide the photons with an effective mass and to allow photon thermalisation at fixed photon number. Under well-chosen experimental conditions\cite{Klaers10a, Kirton13}, despite the continuous drive and dissipation, the photons come into thermal equilibrium with a dye, and the evidence for the Bose-Einstein condensation transition was strong. Due to the room-temperature operation\cite{Plumhof13} and relativity simple experimental set up in comparison to other lower temperature condensates this system offers now an ideal playground to study macroscopic quantum systems.

While Bose-Einstein condensation was initially proposed in the context
of non-interacting Bose gas, the interactions which make this phase
transition experimentally possible lead to a plethora of collective
quantum phenomena. For example, bosons with repulsive interactions,
such as liquid helium-4 and atomic gases, make superfluids while
attractive interactions lead to pairing of fermions and
superconductivity, superfluidity of helium-3, and molecular BECs in
trapped gases.  The interactions in exciton-polariton
condensates\cite{Kasprzak06} play a crucial role in the observed
excitation spectra\cite{Utsunomiya08, Roumpos12} and superfluid
behaviour such as quantised vortices\cite{Lagoudakis08} and persistent
currents\cite{Sanvitto10,Cristofolini13}. Dye-microcavity
  photon BECs share some of the features of exciton-polaritons, such
  as the dissipation, pumping and loss. Unlike exciton-polaritons,
  photon BECs are in the weak light-matter coupling regime, and are
  also thought to be very close to thermal equilibrium. A superfluid
description of light, setting aside the thermalisation process, was
first proposed by R.~Chiao\cite{Chiao99, Bolda01, Tanzini99}, which
follows on from earlier descriptions of inhomogeneous gain media in
optical cavities\cite{Lugiato87, Brambilla91b, Staliunas93}. The
suggestion was to use dilute vapours of alkali metals as a non-linear
medium, together with an optical resonator to constrain the dispersion
relation of the light.

%In the context of the photon BEC, the strength of the photon-photon
%interaction has not yet been accurately measured, and even a full
%theoretical description is still lacking.

In photon BEC, the photon-photon interactions have not yet been accurately measured, and even a full theoretical description is still lacking. Two mechanisms for interactions have been suggested: (i) the Kerr effect of intensity-dependent refractive index, and (ii) a temperature-dependent solvent refractive index, with temperature inhomogeneities driven by inhomogeneities of the light. Preliminary measurements indicated a dimensionless 2D interaction parameter of about $(7\pm 3)\times 10^{-4}$ which is consistent with the second mechanism. However, this mechanism does not act at the single particle level and so does not have any effect on, for example, short-range particle correlations. This thermal mechanism is also slow compared to experimental timescales, typically taking about 1~ms to act, compared to a 1~$\mu$s pump pulse duration. The Kerr effect happens on timescales as fast as the spontaneous emission lifetime of the dye, of the order of 1~ns. The strength of Kerr interactions is unknown, however, expectations that it would be tiny have already motivated theoretical studies on how the phase coherence typical of BECs builds up even in the complete absence of particle-particle interactions\cite{Snoke13}.

In this work, we derive an equation of motion for the pumped photon condensate in a dissipative microcavity, complete with effects of Kerr-type intensity-dependent refractive index, which leads to an effective photon-photon interaction. We use the equivalence of this equation to complex Gross-Pitaevskii equations (cGPEs) used for other systems to help obtain the excitation spectrum and incoherent photoluminescence (IPL), the light that leaks through the cavity mirrors, including the limited experimental resolution. We argue that the IPL spectrum is an excellent diagnostic for the photon-photon interaction strength.  We propose an experimental apparatus for measuring the angle-resolved photoluminescence spectrum to an accuracy sufficient to determine interaction strength even as much as two orders of magnitude more precisely then current estimates. Measurement and understanding of the microscopic origin of interactions in photon BEC are a prerequisite for any possible superfluid effects to be seen in experiments.

\section{Equation of motion for a photon BEC}

The equation of motion for the condensate wavefunction of a photon BEC in a dye-filled microcavity can be derived starting from Maxwell's equations in a non-linear dielectric medium. The closest equivalent theory concerns exciton-polariton condensates, where the light-matter coupling is strong, and therefore the matter component must be treated quantum mechanically. Our Maxwell's equation approach is valid only for photon BEC, where the light-matter coupling is weak. The cavity is so short that only one longitudinal mode is relevant.

We propagate the wave from one mirror of the cavity to the other and back again. The net change in electric field over one cycle of propagation divided by the time that cycle takes determines the time derivative for the electric field. Note, that it is also possible to obtain a similar equation of motion by considering a decomposition over quasi-normal modes of the optical resonator in appropriate paraxial and slowly-varying-envelope approximations \cite{Lugiato88}.

\subsection{Non-linear wave propagation}

\begin{figure}[hbt]
	\centering
	\includegraphics[width=0.8\columnwidth]{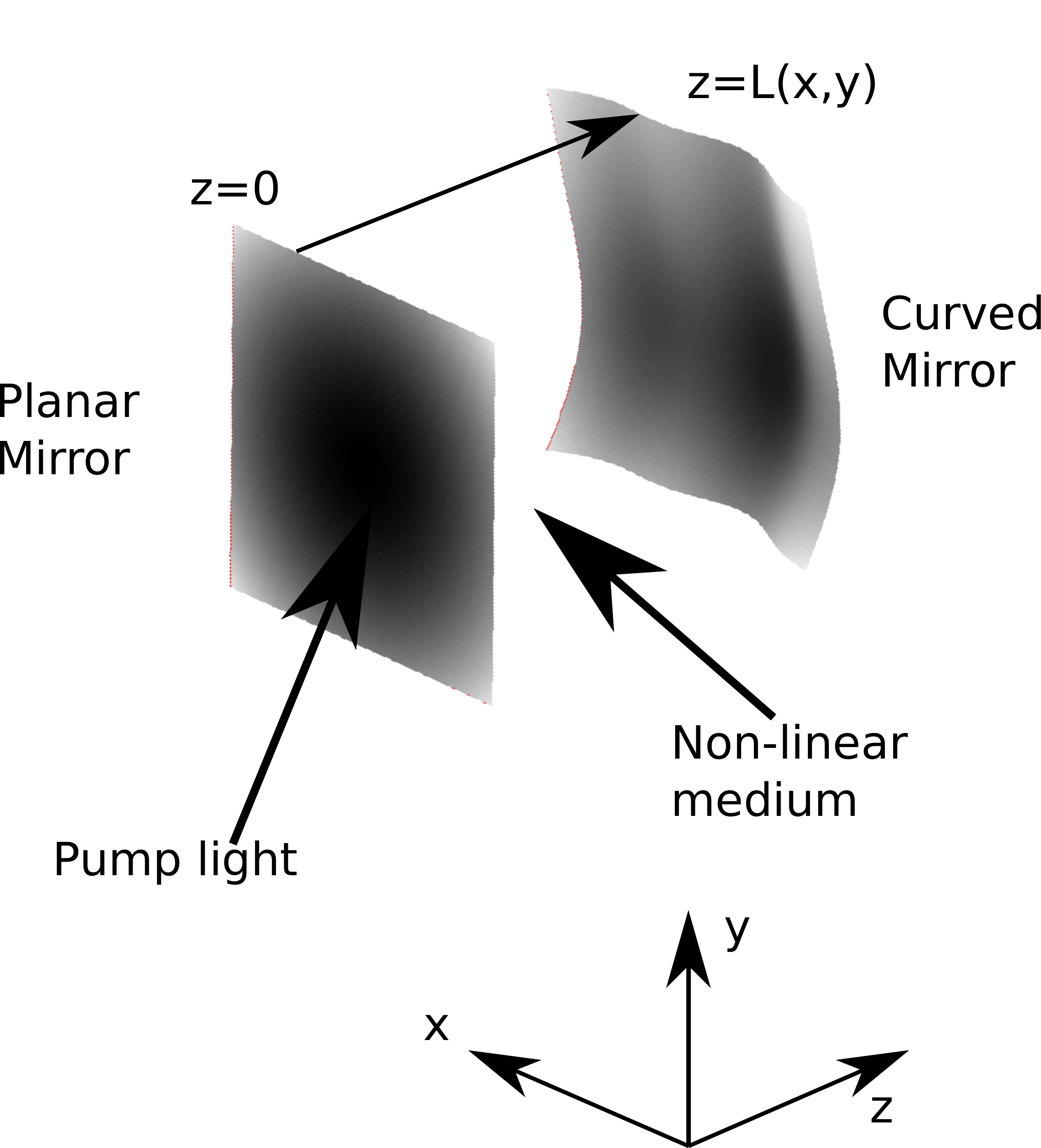}
	\caption{A cavity filled with a non-linear medium which allows
          the photons to come to thermal equilibrium faster than they
          leave the cavity. The co-ordinate system is as used in the
          rest of this article. The length variations of the cavity
          play the role of a potential energy landscape for the
          photons. The cavity is so short that only one longitudinal
          mode is excited, and individual photons have a dispersion
          relation equivalent to that of massive particles.}
	\label{Fig: define coords}
\end{figure}

The non-linear electric polarisability of the medium can be accounted
for by writing the electric polarisation as a linear part plus a
non-linear part\cite{Boyd}: \mbox{${\bf P} = {\bf P}_L + {\bf
    P}_{NL}$}. The constitutive relation is \mbox{${\bf D} =
  \epsilon_0 \epsilon_L {\bf E} + {\bf P}_{NL}$}, where the linear
permittivity (in the limit of low intensity light) is $\epsilon_L =
n_L^2 = 1+\chi_L$ and $\chi_L$ is the linear susceptibility. We define
$n_L$ as the refractive index at low intensity, and the wave equation
for the electric field becomes:
\begin{align}
		\nabla (\nabla \cdot {\bf E}) - \nabla^2 {\bf E} +
	\frac{n_L^2}{c^2}\ddot{\bf E} =-\frac{1}{c^2}\ddot{\bf P}_{NL}
	\label{Eqn: full wave eqn}
\end{align}
with the dot representing the time derivative. We make the paraxial
approximation and assume that the electric field can be written as a
scalar (constant polarisation throughout space, perpendicular to the
direction of propagation). We consider a travelling wave solution,
$E=E_0(x,y,z)\ei{(k_L z- \omega t)}$ with a slowly-varying envelope:
$z$ is the axis of the optical resonator, as shown in \figref{Fig: define coords}. 

For annotation's sake, where only one argument of $E_0$ is given, it is $z$; $x$ and $y$ are left
implicit. With two arguments, they are $x$ and $y$, leaving $z=0$
implicit. The angular frequency and wavenumber of the light are
$\omega$ and $k_L = \omega n_L/ c$ respectively.  Now, using
conventional definitions for the Kerr-type non-linearity, the paraxial
wave equation becomes:
\begin{align}
	-\left\{2{\rm i}k\partialderiv{E_0}{z} +
	\partialderiv{^2E_0}{z^2} + \nabla_\perp^2E_0\right\} =
	\frac{k_L^2}{n_L^2} 3 \chi^{(3)}|E_0|^2 E_0 \label{Eqn:
	paraxial wave eqn}
\end{align}
where $\nabla_\perp^2 = \partialderiv{^2}{x^2}
+\partialderiv{^2}{y^2}$. Starting from the electric field at $z=0$,
we can find the field some small propagation distance away at $z=L$
using a first-order Taylor expansion:
\begin{align}
	E_0(L) \simeq E_0(0) -
	\label{Eqn: nonlin wave propagation} \frac{L}{2{\rm i}k_L} \left\{ \nabla_\perp^2 E_0(0) 
	 + \frac{k_L^2}{n_L^2} 3 \chi^{(3)}|E_0(0)|^2 E_0(0)\right\}.
\end{align}

\subsection{To and fro between mirrors}

We now consider a forward travelling wave with an electric field envelope \mbox{$E_0^\rightarrow(x,y,0)$} and use \eqnref{Eqn: nonlin wave propagation} to find the electric field at the other mirror. We will allow the mirror surface position to vary, so the cavity length is a function of position: $L(x,y)=L_0+\delta L(x,y)$. For length variations which are much less than the wavelength of the light, we make a linear approximation for the phase of the light, and neglect additional envelope propagation effects. The forward propagating field at the second mirror is:
\begin{align}
	E&_0^\rightarrow(x,y,L(x,y))= E_0^\rightarrow(0)\left(\eim{k_L
		L_0}+{\rm i}k_L\delta L \right)\\ &-\frac{L_0}{2{\rm
		i}k_L} \ei{k_L L_0}\!\left\{ \nabla_\perp^2
		E_0^\rightarrow(0) + \frac{k_L^2 }{n_L^2} 3
		\chi^{(3)}|E_0^\rightarrow(0)|^2
		E_0^\rightarrow(0)\right\}. \nonumber
\end{align}
The mirrors have transmission and reflectance $\tau$ and $r$ respectively. After reflection, the backwards propagating field at $L(x,y)$ is $E_0^\leftarrow(L(x,y)) = -rE_0^\rightarrow(L(x,y))$. Propagation to the first mirror at $z=0$ and subsequent reflection follow the same pattern. Finally, to complete a cycle, inhomogeneous pump light enters through the mirror at $z=0$. In order to model saturable, incoherent pumping, it is sufficient to write the pump term in the form $\left(\alpha - \beta|E_0|^2\right)E_0$. Here, $\alpha$ represents difference between gain via stimulated scattering of photons into the condensate and cavity mirror loss, and $\beta$ governs saturation\cite{Keeling08}. 

The time taken for a cycle is given by $\delta t = c / 2n_L L_0$, where $c$ is the speed of light in vacuum. The electric field change $\delta E_0$ in one cycle is derived as above, and then we convert finite differences to derivatives: \mbox{$\delta E_0 / \delta t = \partialderiv{E_0}{t} = \dot{E}_0$}. High-order terms in small quantities are neglected. The equation of motion for the electric field envelope at the first mirror, assuming that it varies slowly compared to the cavity round-trip time becomes:
\begin{align} \frac{1}{{\rm i}\,\omega}\partialderiv{E_0}{t} = &
	\left[
		\left(\frac{\delta L}{L_0} - \frac{\delta\omega}{\omega} \right) 
		+ \frac{1}{k_L^2}\nabla_\perp^2
		+ \frac{3\chi^{(3)}}{2n_L^2}|E_0|^2
	\right]E_0 \nonumber\\
	&
	- \frac{\rm i}{2q\pi}\left[ 
		\left(\alpha - \beta\left|E_0\right|^2\right)- (1-r) \right]
	E_0,\label{Eqn: of motion}
\end{align}
where $\chi^{(3)}$ is the Kerr non-linearity; $\delta \omega$ the cavity detuning; $k_L = 2\pi q/L_0 = \omega n_L / c$; $n_L$ the refractive index in the limit of low light intensity; $q$ the longitudinal mode number. Only the lowest relevant order in small quantities is retained. This equation describes only the condensed photons, and not the thermal (non-condensed) particles.

Finally, it is straightforward to include the effects of inhomogeneous
linear refractive index (which appears as an increase in the effective
length of the cavity), or spatially variable mirror reflectivities.

\subsection{Similarity to a complex GP equation}
A typical form for the cGPE for the condensate wavefunction
$\psi$ in a system with incoherent pumping (as in photon BEC) is:
\begin{align}
  -{\rm i}\hbar\partialderiv{\psi}{t} &= \label{Eqn: cGPE} %\\
  &\hspace{-2ex}\left[ V({\bf r})\!-\!\frac{\hbar^2}{2 m}\nabla^2_\perp 
    + g|\psi|^2 + {\rm i}\!\left(\gamma_{net} -\!
      \Gamma|\psi|^2\right) \right]\psi
  %\nonumber
\end{align}
where $\gamma_{net}$ is the difference between the pump rate and cavity decay rate and is equal to $\alpha$ in \eqnref{Eqn: of motion}, and $\Gamma$ describes the saturation of pumping (ensuring stability). In the steady state  \mbox{$\Gamma = \gamma_{net} / |\psi(0,0)|^2$}. In two dimensions $\int \drr\, |\psi(x,y)|^2 = N_{BEC}$, where $N_{BEC}$ is the number of particles in the condensate.

Comparing (\ref{Eqn: of motion}) and (\ref{Eqn: cGPE}) the first term
in Eqn. (\ref{Eqn: of motion}) is equivalent to potential energy in
the cGPE with an additional energy offset due to the
detuning between the cavity mode and the light. The second term comes
from diffraction of the light, and corresponds to kinetic energy,
while the third term to the interactions. The energy stored in the
electric field of the standing wave in the cavity is
\mbox{$\frac{1}{2} n_L^2 L_0 \epsilon_0\int\drr |E_0(x,y)|^2 =
  N_{BEC}\hbar\omega$}. We can define the quantity $m$ through $\hbar
\omega = m c^2 / n_L^2$, and this will play the role of an effective
photon mass. With these analogies, we can convert the equation of
motion for $E_0$ electric field \eqnref{Eqn: of motion} into the
cGPE \eqnref{Eqn: cGPE} with the steady-state mean-field
solution $\psi = \psi_0 {\rm e}^{-{\rm i}\mu t / \hbar}$ where $\mu$ is
the chemical potential:
\begin{align}
	\psi_0 &= E_0 \sqrt{\frac{n_L^2 \epsilon_0 L_0}{2 \hbar
	\omega}}\\ g&= \frac{3\hbar^2\omega^2}{n^4_L \epsilon_0
	L_0}\chi^{(3)} = \frac{\hbar^2}{m} \tilde{g}
	\label{Eqn: gtilde}
\end{align}
where $\tilde{g}$ is the dimensionless 2D interaction parameter\cite{Hadzibabic11}. 

\section{Excitation spectrum and photoluminescence}

The IPL, which can be measured by angle- and
energy-resolved techniques, for a system in thermal equilibrium is
given by the Bose-Einstein occupation function $n_B(\omega)$ times the
spectral weight\cite{Marchetti07,Szymanska07}:
\begin{align}
	P_L(\kom) = n_B(\omega) W(\kom).
\end{align}
The spectral weight can be obtained from the retarded Green's function
for the response of the system to perturbations, $G_R$:
\begin{align}
	W(\kom) =  2\,{\tt Im}\left[G_R(\kom)\right],
\end{align}
as shown for example in Ref.~\cite{Keeling05}. The exact form of $G_R$
depends on the model used and here we determine $G_R$ for a
dissipative and driven case given by \eqnref{Eqn: cGPE} (the open
system model) as well as for a simplified case where the dissipative
terms are neglected (the closed system model). The two models give
almost identical IPL since the decay processes in a photon BEC are
very small. Thus we further use the closed system model to determine
the influence of the trapping potential on IPL.

\subsection{Open-system model, to deal with dissipation}

In the homogeneous case, $V(x,y)=0$, the steady-state mean-field
solution of \eqnref{Eqn: cGPE} $\psi_0$, is found, by writing a
solution with time variation $\eim{\mu t / \hbar}$ to obtain the
chemical potential \mbox{$\mu = g|\psi_0|^2$}. Note that in the
Thomas-Fermi-limit of a harmonic trap with frequency $\Omega_0$, the
chemical potential is $\hbar\,\Omega_0 \sqrt{\frac{\tilde{g}
    \,\,N_{BEC} }{\pi}}$. The equation of motion for small (linear)
variations about this mean field, \dpsi\ can be obtained by starting
from a total wavefunction as $\annihilation = \psi_0+\dpsi$ and
subtracting the mean-field solution.

Comparison with the Hermitian conjugate leads to a system of linear
equations in \dpsi\ and \dpsidag. The matrix operator which relates
the two is the inverse Green's function\cite{Marchetti07},
$(\mathcal{G})^{-1}$, which is also known as the Bogoliubov
operator\cite{Carusotto13}:
\begin{align}
	-{\rm i}\hbar\partialderiv{}{t}\begin{pmatrix}\dpsi \\
          \dpsidag\end{pmatrix}  
		= 
	\left(\mathcal{G}\right)^{-1}\begin{pmatrix}\dpsi \\
          \dpsidag\end{pmatrix}. 
\end{align}
The relevant component for the photoluminescence is the diagonal ($\delta\annihilation$, $\delta\annihilation$) component of the full Green's function: \mbox{$G_R(\kom) = \mathcal{G}^{11}(\kom)$}.

In the stationary and homogeneous case, the inversion is most easily
performed in Fourier space to give \cite{Roumpos12}:
\begin{align}
	\mathcal{G_R}(\kom) = &\,\frac{1}{\hbar^2\omega(\omega + 2{\rm
		i}\gamma_{net}) \,-\, \ek(\ek + 2\mu)} \,\,\times
	\label{Eqn: Green's function}\\
	&\begin{pmatrix}
		\mu + \ek + \hbar\omega + {\rm i}\hbar\gamma_{net} 
			& -\mu + {\rm i}\hbar\gamma_{net}\\
		-\mu - {\rm i}\hbar\gamma_{net} 
			&  \mu + \ek - \hbar\omega + {\rm i}\hbar\gamma_{net}
	\end{pmatrix}
	\nonumber
\end{align}
where $\ek = \frac{\hbar^2k^2}{2m}$ is the kinetic energy of a free
particle, and ${\bf k}$ is in the $x,y$ plane. 

The IPL from the dye-microcavity
sample at temperature $T$ using the open system model of \eqnref{Eqn:
  cGPE} but assuming that photons thermalise is
\begin{align}
	P_L^{(open)}(\kom) = &\frac{4\hbar}{e^{\frac{\hbar\omega}{ k_B
            T}}-1}
%\times \label{Eqn: PL open}
%\\ &  
\frac
          {\gamma_{\rm{net}}(\ek+\omega)(\ek+2\mu+\omega) }
          {4\gamma_{\rm{net}}^2\omega^2+(\ek^2+2\ek \mu -\omega^2)^2}.  \nonumber
\end{align}
The energy scale is relative to the chemical potential, so $\mu$ does
not appear in distribution factor. We note that this expression is approximate. We have assumed a thermal distribution of fluctuations (non-condensed photons). For full self-consistency, a frequency-dependent  distribution of fluctuations would imply frequency-dependent
decay and pump rates, which, for simplicity, we have approximated as being frequency independent. However, only small corrections in the relevant region around the poles would come about from considering non-Markovian decay rates.

\begin{figure}[hbt]
	\centering
	\includegraphics[width=0.8\columnwidth]{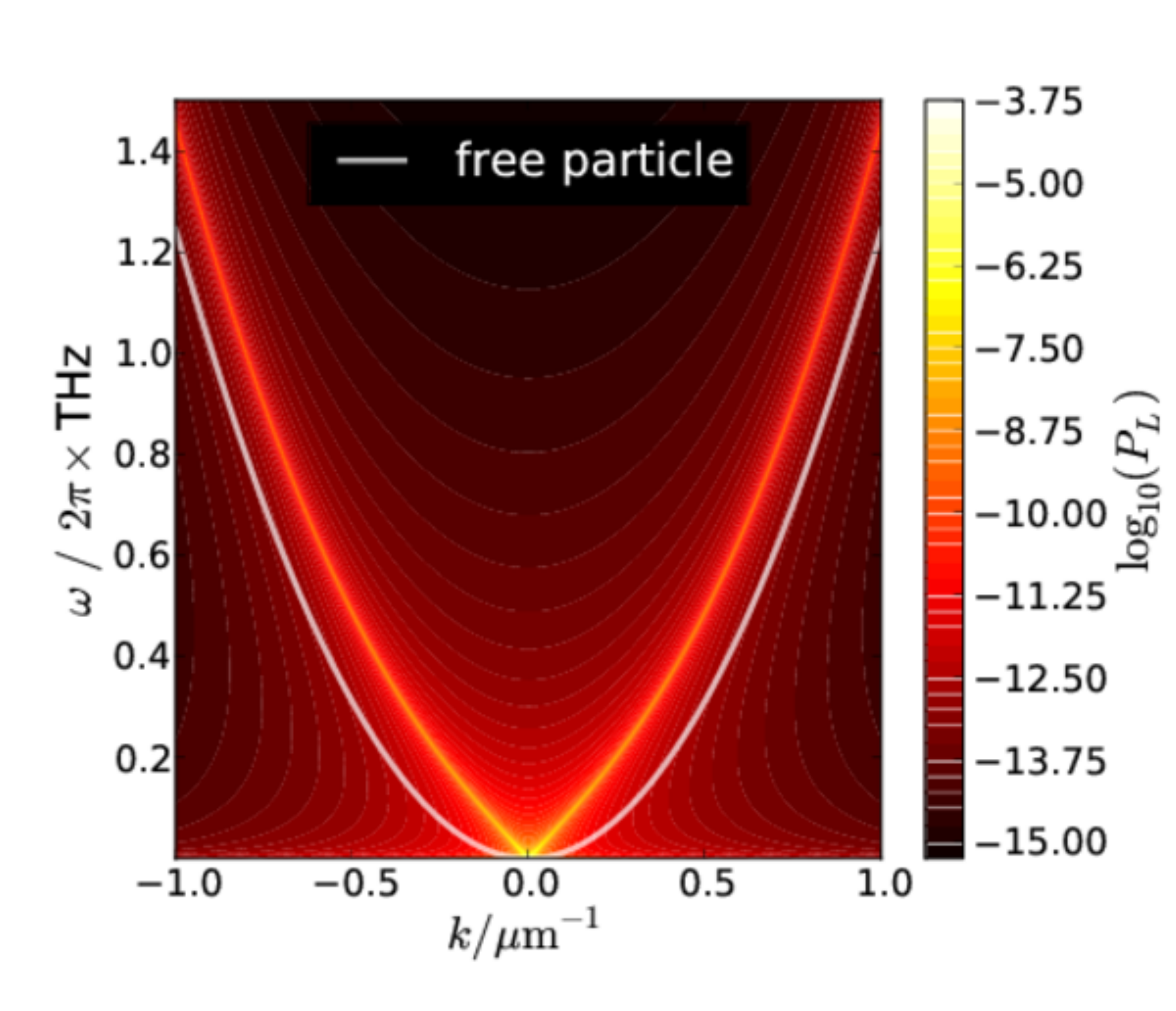}
	\includegraphics[width=0.8\columnwidth]{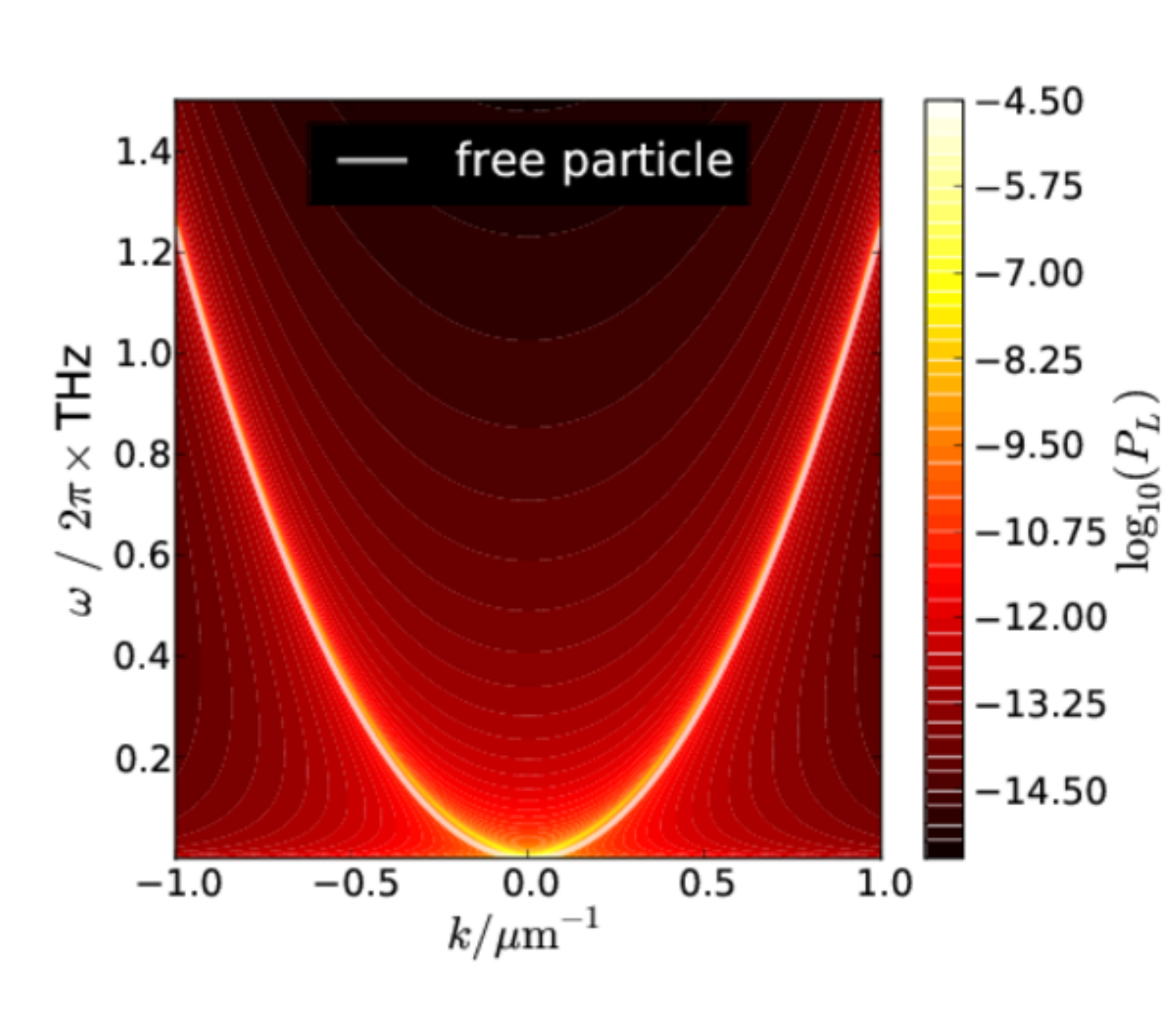}
	\caption{Examples of photoluminescence, calculated using the
          open-system model. On this scale, the closed system results
          look the same. Top: dimensionless interaction parameter
          $\tilde{g}=10^{-3}$ with $10^5$ condensed photons. Bottom:
          $\tilde{g}=10^{-5}$ and $10^5$ condensed photons. As a guide
          for eye we plot the dispersion relation for free particle
          with the same photon mass. Angular frequency $\omega$ is
          taken relative to the chemical potential. Other parameters:
          $\gamma_{net} = 2\pi\times$1~GHz; $T=300$~K; central density
          calculated from Thomas-Fermi profile of photons in a
          circularly-symmetric harmonic trap of frequency 40~GHz.}
	\label{Fig: PL}
\end{figure}

\figref{Fig: PL}\ shows the IPL energy-momentum spectrum, in which the
Bogoliubov dispersion is clearly visible. The parameters of the
calculation are experimentally achievable, and similar to values in
Ref.~\cite{Klaers10b}. For values of $\tilde{g}$ above about
$10^{-5}$, the difference between free particles and photon
quasiparticle excitations from the interacting condensate is very
clear.

\subsection{Closed-system model, to deal with inhomogeneities}

By ignoring the dissipative term in \eqnref{Eqn: cGPE} we
determine the IPL for the closed system. In the homogeneous case, the
difference between the open- and closed-system models is mostly
notable at low momentum and energy. The closed system follows the
usual Bogoliubov modes down to zero momentum but the open system modes
become diffusive at very low momenta\cite{Szymanska06,Wouters07,
  Carusotto13}. However, the effects of interactions are observable in
the IPL energy-momentum spectrum at moderate momenta and energy, and
so the two models largely agree for the purposes of this work.

Photon BEC, however, is not homogeneous.  In the local density
approximation (LDA), we proceed by using a local chemical potential
with \mbox{$\mu'({\bf r}) =\mu - V({\bf r})$}. The energy spectrum for
excitations is \mbox{$\xik({\bf r}) = \sqrt{\ek(\ek + 2\mu'({\bf
      r}))}$}. This is a local version of the Bogoliubov
spectrum\cite{Fedichev98}. There are finite temperature
corrections\cite{PitaevskiiStringari} which we are neglecting
here. The local spectral weight is then given by:
\begin{align}
	W^{(closed)}(\kom; {\bf r}) \!=& \frac{\ek \!+\! \mu'({\bf r})
		\!+\! \xik({\bf r})}{2\xik({\bf
		r})}\,\delta(\hbar\omega\!-\!\xik({\bf r})) \\
		&\hspace{0ex}- \frac{\ek + \mu'({\bf r}) - \xik({\bf
		r})}{2\xik({\bf r})}\,\delta(\hbar\omega+\xik({\bf
		r})).\nonumber
\end{align}
The energy scale is shifted such that excitations of energy
$\hbar\omega=0$ are at the chemical potential. Some broadening is put
into the system by hand by adding an imaginary part $\hbar\kappa$ to
the energy, $\epsilon \rightarrow \epsilon -{\rm i}\hbar\kappa$ and,
when integrated around the zero of the argument, $\delta(\epsilon)
\rightarrow
\frac{1}{\pi}\,\frac{\hbar\kappa}{\epsilon^2+\hbar^2\kappa^2}$. The
local Bose occupation factor becomes \mbox{ $n_B(\omega; {\bf r}) =
  \frac{\hbar}{e^{[\hbar\omega - V({\bf r})] / k_B T}-1} $} where
\mbox{$n_B(\omega) = \iint {\rm d}^2{\bf r}\, n_B(\omega; {\bf r}) /
  A$} and $A$ is a typical area of the system, e.g. $2\pi \mu / m
\Omega_0^2$ for in the Thomas-Fermi limit in a harmonic potential. The
total photoluminescence observable becomes:
\begin{align}
	P_L^{(closed)}(\kom) = \iint {\rm d}^2{\bf r} \, n_B(\omega; {\bf
	r})W^{(closed)}(\kom; {\bf r}).
\end{align}
An inhomogeneous confining potential means that the density varies
across the condensate, which in turn leads to variations in the
Bogoliubov spectrum, i.e. the speed of sound. The integration over all
positions of the condensate causes the lines in the IPL
energy-momentum spectrum to broaden, making it more difficult to see
the effects of interactions. The photoluminescence calculated here is
the incoherent part; coherent photoluminescence will be emitted from
the condensate mode. It may contain a large range of momenta, but it
will all be at the lowest energy available, and so incoherent and
coherent light can easily be distinguished (condensate broadening in
energy is expected to be very small on the scale of \figref{Fig:
  PL})

\section{Energy-momentum spectroscopy}

\begin{figure}[tbh]
	\centering
        \includegraphics[width=0.9\columnwidth]{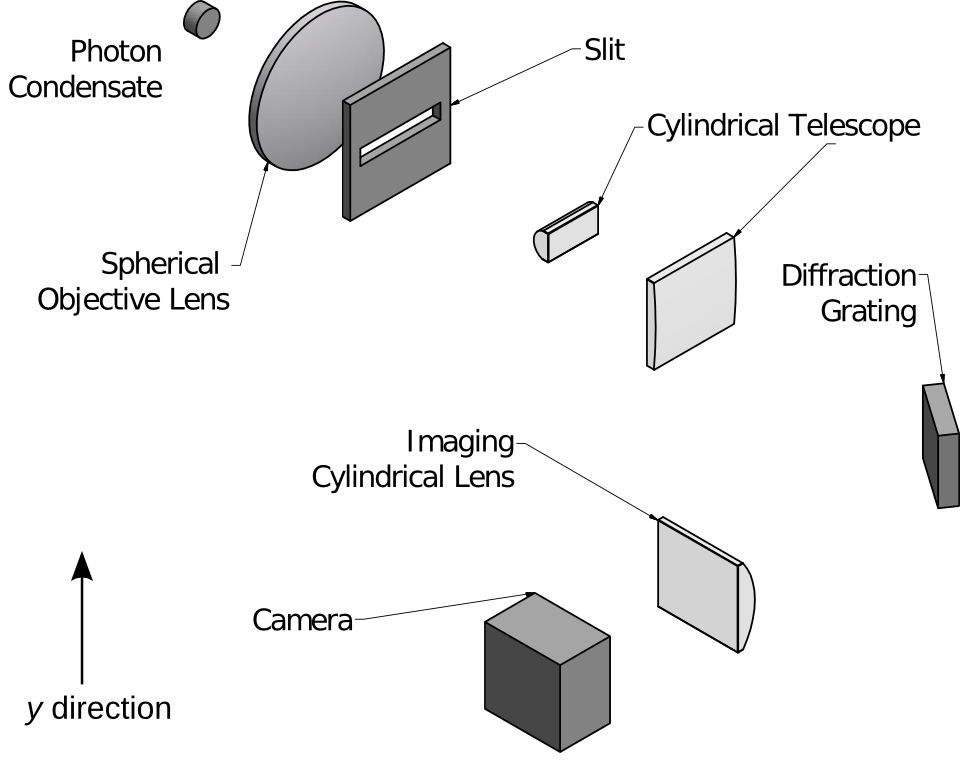}
	\caption{A schematic diagram of Angle-Resolved
            Photo-Luminescence Spectroscopy
            (ARPLS). Photoluminescence from the photon condensate is
          collimated by the (spherical) objective lens. Components
          with wavenumbers $k_y \simeq 0$ only pass through the
          slit. A cylindrical telescope magnifies the light in the
          $y$-direction, before it strikes a reflective diffraction
          grating, whose rules run parallel to the $x$-direction. The
          (cylindrical) imaging lens then ensures that the image on
          the camera corresponds to momentum ($x$) and energy ($y$).}
	\label{Fig: ARPLS diagram}
\end{figure}

It is possible to observe the photoluminescence resolved in both
energy and one component of momentum. The angle-resolved
  photo-luminescence spectrum (ARPLS) of exciton-polariton samples
has been successfully measured, demonstrating the effect of
interactions on the Bogoliubov dispersion relation\cite{Utsunomiya08,
  Roumpos12}. The basic experimental optical apparatus to be used for
photon BEC ARPLS is shown in \figref{Fig: ARPLS diagram}.

The angle of a photon relative to the optic axis inside the cavity is
$\boldsymbol{\theta}_{int} = (\theta_x, \theta_y) = \arctan({\bf k} /
k_0)$ with $k_0 = q\pi n_L/ L_0$ being the typical wavenumber. The
angle that the emitted light makes to the optic axis is then
$\boldsymbol{\theta}_{ext} = \arcsin{( n_L \sin
  \boldsymbol{\theta}_{int})} $. The photoluminescence is at the focus
of the objective lens of focal length $f_{obj}$, and so the
displacement from the optic axis is $(x,y) = n_L f_{obj} {\bf k} /
k_0$ (in the small angle approximation). Behind the objective, the
light passes through a slit of size $d_{slit}$ in the $y$ direction so
only the $k_y\simeq 0$ components make it through. The image in
momentum space is unaffected by subsequent optics. In order to improve
spectral resolution, the image is magnified in the $y$ direction by a
cylindrical telescope of magnification $M_y$ and it strikes a
reflective grating, whose lines are parallel to the $x$ direction and
spaced by $d_{grating}$ in $y$. The first-order diffraction angle in
the $y$ direction is $\theta_y = \arcsin(\lambda/d_{grating})$, with
$\lambda$ being the wavelength of light. The light passes through a
cylindrical imaging lens of focal length $f_{im}$, and the camera sits
at this focus.

\subsection{Experimental limits to measuring interactions}

The mapping between \kxlam\ and position on screen is blurred by diffraction in the propagation of the light from source to detector. We analyse what the minimum resolvable momentum and energy would be, and what that means for the minimum resolvable interaction strength.  In general, interactions will be detectable when the Bogoliubov spectrum is significantly different from the free-particle spectrum. This happens on energy scales less than $\xi_{min} \simeq 2\mu$, and momentum scales less than \mbox{$p_{min}\simeq 2\sqrt{m \mu}$}.

For numerical evaluation, we need a value of the dissipation rate $\gamma_{net}$ of \eqnref{Eqn: cGPE}. The rate of scattering into the condensate is of the same order as the rate photons scatter from the dye, which depends on the dye concentration: $\gamma_{R} = 1 / n_{dye}\sigma_{dye}(c/n_L)$ where $n_{dye} \simeq 10^{24}$~molecules/m$^3$, and is typically about \mbox{$2\pi\times $2--6~GHz} for Ref. \cite{Klaers10a, Klaers10b}. An experimentally achievable cavity loss rate, $\kappa_{cav} \simeq 2\pi\times 1$~GHz, and is governed by the mirror quality. Then, $\gamma_{net} = \gamma_R - \kappa_{cav}$ is of order $2\pi\times 1$~GHz.

\subsubsection{Resolution in momentum}

The monochromator optics for energy resolution means that the
Fourier-space image (after the objective lens and slit) will propagate
and diffract before it reaches the camera. The propagation distance
between objective lens and camera is $L_{prop}$. A range of small
transverse wavenumbers $\delta_k$ corresponds to a region of size $n_L
f_{obj} \delta_k / k_0 $ at the objective, which will diffract to a
region of size \mbox{$\delta_x^{(cam)} = 2 L_{prop} / n_L f_{obj}
  \delta_k$} at the camera (in the far field). Inverting this
expression gives the diffraction limit for transverse
wavenumber. Considering the mapping between angle and position, the
equivalent limit set by the pixel size $\delta_x^{px}$ of the camera
is $\delta_k/k_0 = \delta_x^{(px)} / n_L f_{obj}$. The optimum results
will be achieved with diffraction limit roughly equal to pixellisation
limit, and we find:
\begin{align}
	\delta_k^{(min)} = \frac{2}{n_L f_{obj}}\sqrt{\frac{\pi L_{prop}}{\lambda}}.
\end{align}
The same analysis yields an optimal pixel size of
\mbox{$\delta_x^{(px)}=\sqrt{L_{prop}\lambda / \pi}$}. Putting in
plausible experimental values $f_{obj}=0.2$~m, $L_{prop}=0.3$~m and
$\lambda = 580$~nm, we obtain a minimum in-plane momentum resolution
of $1.3\times10^{4}$~m$^{-1}$. The appropriate camera pixel size would
be about 180~$\mu$m, which is trivially achievable.
It is clearly advantageous to use a long focal length objective: to
collimate a useful range of momenta, a lens with $f_{obj}=0.2$~m
should be 20~mm diameter. The equivalent size of the slit in momentum
space should be no bigger than the expected momentum resolution of the
entire optical system. In real space, that means that the slit should
be about the same size as the detector pixels, $d_{slit} \sim
180$~$\mu$m.

\subsubsection{Resolution in energy}

Energy resolution is limited by the size of the beam at the
grating. The resolving power for the first-order diffraction fringe is
approximately equal to the number of grating lines covered by the
incident beam: \mbox{$\delta_\lambda= \lambda d_{grating} / D$}.
Reasonable experimental parameters are
\mbox{$1/d_{grating}=900$~lines$/$mm} and $D=d_{slit}M_y = 18$~mm
(implying a cylindrical telescope of magnification approximately
75). The resulting wavelength resolution is $\delta_\lambda =
0.04$~nm, or equivalently $\delta_\epsilon = h\times$30~GHz for 580~nm
emission. With an imaging lens focal length of $f_{im}=50$~mm, the
detector pixel size required not to compromise this resolution is
4~$\mu$m: a commonplace pixel size for a CCD camera.

\subsection{Minimum detectable interaction strength}

For reasonable parameters of \mbox{$\Omega_0 = 2\pi\times$40~GHz} and
\mbox{$N_{BEC}=10^5$}, and optics as previously described, we find that it
should be possible to resolve interactions as weak as \mbox{$\tilde{g}_{min}
\simeq \left(\frac{\hbar^2\delta_k^2}{4m}\right)^2\frac{\pi
}{N_{BEC}(\hbar\Omega_0)^2} = 2\times 10^{-10}$} (if momentum is the
limiting resolution) or \mbox{$\tilde{g}_{min} \simeq
\delta_\epsilon^2\frac{\pi }{N_{BEC}(2\hbar\Omega_0)^2} = 2\times
10^{-5}$} (if energy resolution is the limiting factor). It is worth
noting that for extremely weak interactions, the Thomas-Fermi
approximation used in deriving the photoluminescence spectrum is
unlikely to be very accurate.

The implication for the experimenter is that the momentum is easily
resolved, so most experimental effort will be required to attain the
best possible energy resolution. Interactions 40 times weaker than
those reported in Ref.~\cite{Klaers10b} should be detectable. Advanced
data analysis could further improve the sensitivity. An example of
plausible experimental data is shown in \figref{Fig: ARPLS diagram
realistic}, which includes both energy and momentum instrumental
broadening and the effects of saturation and finite dynamic range of
the detector camera (noise is not included in the model).

\begin{figure}[hbt]
	\centering
        \includegraphics[width=0.8\columnwidth]{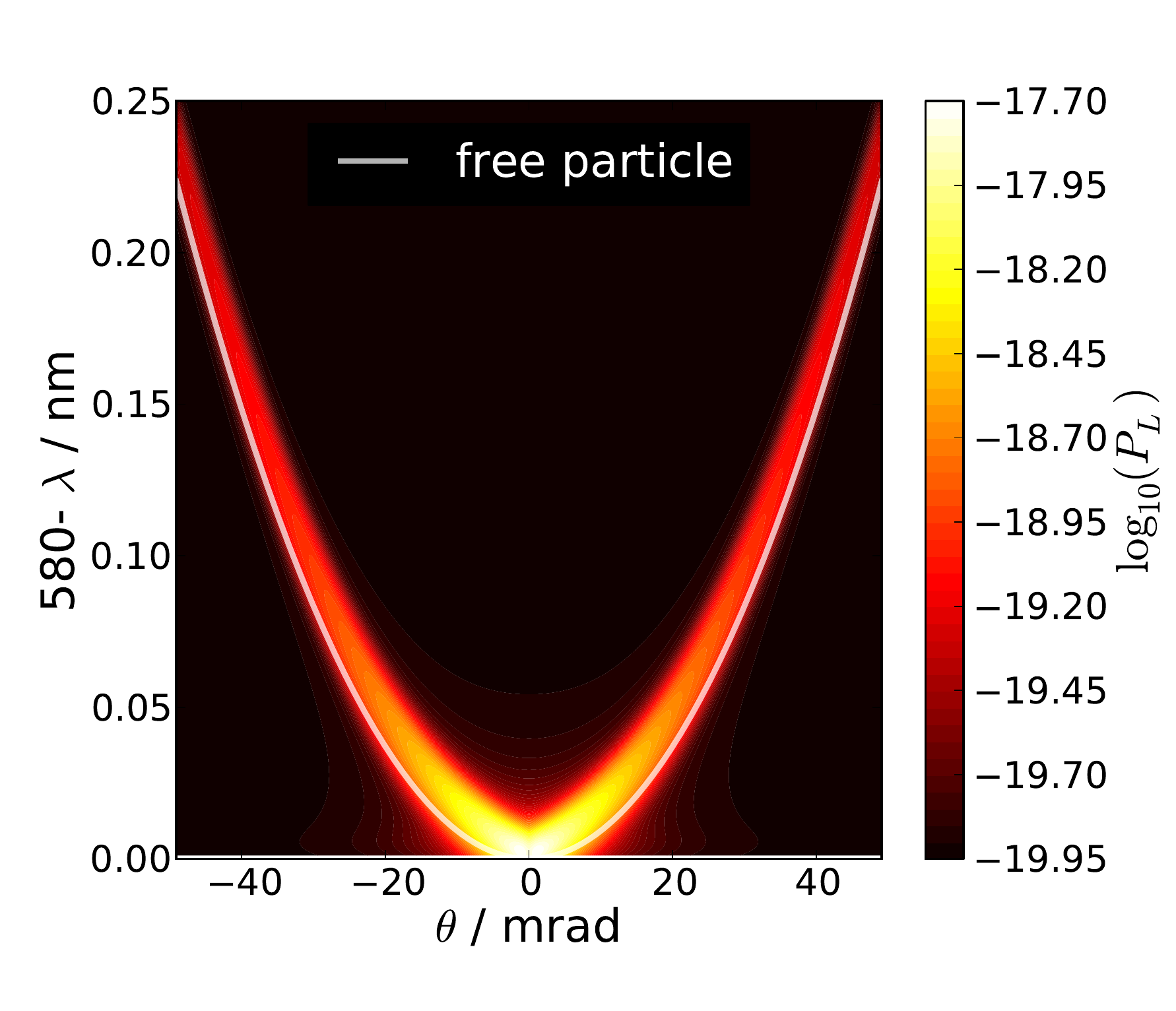}
	\caption{Observable photoluminescence energy-momentum spectrum
          using a closed-system model, including the effects of the
          inhomogeneous confining potential, finite instrumental
          resolution in momentum and energy, and the finite dynamic
          range of a typical camera. Experimental parameters are
          described in the main text. Dimensionless interaction
          strength, number of condensed photons, temperature and trap
          frequency are $10^{-5}, 10^5, 300$~K and $40$~GHz
          respectively.}
	\label{Fig: ARPLS diagram realistic}
\end{figure}

If the energy resolution required cannot be matched by grating
spectroscopy, then an external Fabry-Perot cavity spectrometer would
be a viable alternative. The minimum energy resolution in that case would
most likely be set by the intrinsic linewidth of the resonator which
contains the photon condensate, probably about 1~GHz. The limit on
$\tilde{g}$ would in that case be somewhere around $10^{-6}$.

There is very little literature on the non-linear susceptibility of
dyes like Rhodamine. The closest available data is for very short
pulses, which gives an underestimate of the non-linearity since the
steady-state excited state population has not been reached. Taking a
reasonable value for the scattering cross-section of
Rhodamine\cite{Schaefer} of $\sigma = 2\times 10^{-22}$~m$^2$, and
using the result of Ref.~\cite{Delysse98}, we infer $\chi^{(3)} \simeq
5\times 10^{-20}$~(m/V)$^2$. This in turn implies a lower bound for
the 2D dimensionless interaction parameter: $\tilde{g} > 2\times
10^{-7}$.

The intensity-dependent refractive index may come from an effect as
simple as saturation of the excited state population. For two-level
systems, at short wavelengths one expects negative $\chi^{(3)}$
leading to repulsive interactions, but attractive interactions for
long wavelengths. This frequency-dependent (i.e. also time-dependent)
interaction strength leaves open the possibility for retarded
interactions which will complicate the analysis of excitations about
the condensate, and could lead to so-far unpredicted phenomena.

\section{Conclusions}

To conclude, we believe that the photoluminescence spectrum from a photon BEC
can be observed using standard optical elements (lenses, a diffraction grating
and a camera), with a sufficient resolution to detect dimensionless interaction
parameters as small as about $10^{-5}$. The method measures only the fast,
Kerr-type interactions and avoids the apparent interactions which come from the
temperature-dependent refractive index of the solvent. If a Fabry-Perot
resonator were used, the resolution may be an order of magnitude better. This
compares well to the best available data in the literature on non-linear
susceptibilities in Rhodamine dyes, and we can expect interaction effects in
photon BECs to be experimentally observed via the energy-momentum spectrum.
Knowledge of the magnitude and nature of the interactions in photon BEC is
important for the observation of photon superfluidity. Likewise, applications in
quantum metrology, i.e. optical measurement of fragile samples, depend on the
photon-photon correlations, which are strongly affected by the microscopic
nature of the interactions.

We are grateful to EPSRC for funding this work (RAN for the fellowship
EP/J017027/1 and MHS under fellowship EP/K003623/1 and grant
EP/I028900/1). We thank Jan Klaers, Jonathan Keeling and Peter Kirton
for informative discussions.\\

\tobeignored{
\section{Popular summary}

%<250 words, 
%first paragraph of the Summary should be intended for general readers, ...1) giving the most pertinent general context; 2) posing the question(s) or problems or describing the issues that the underlying paper addresses and explaining why the question(s) is(are) worth asking; 3) describing succinctly (preferably in one sentence) what the paper does (or achieves) with regard to the questions. 
%
% second paragraph (and a third, if necessary) should provide a relatively more extensive description of the actual content of the work. Th
%
%final paragraph may give an outlook to new research t

Thermal equilibrium happens when the system under consideration is one of the most probable states available given its interaction with a reference, e.g. a fixed temperature bath, which controls some parameters, e.g. temperature. When the particles of the system are identical, quantum statistics defines which states are most probable. Particles of light, photons, are bosons, so at low temperature they ought to form a Bose-Einstein condensate (BEC). However, the number of photons is not conserved unless the photons are coupled to matter in some way, so condensation is not normally seen unless the thermalisation conserves number, which can be done with fluorescent dye in a microcavity, even at room temperature. In this article we derive the equation for photon Bose-Einstein condensate and show how some of the microscopic properties affect observable quantities.

We derive the equation, which looks like the equation for BEC of either exciton-polaritons or trapped atoms. However, our equation does not start from matter and then build in the light, but starts from Maxwell's equations for the light itself. We then study the fluctuations around the mean-field which show up in the spectrum of excitations. That spectrum could then be used to infer if and how the photons interact with each other. 

Photon-photon interactions are weak, even in dye, but might be strong enough to lead to superfluid behaviour in photon BECs. Light behaving as a fluid would be an unusual and exciting thing.
}

\bibliographystyle{prsty}
\bibliography{pbec_interactions_refs}

\end{document}